\documentclass[prx,aps,twocolumn,floatfix,citeautoscript,showpacs]{revtex4-2}
\usepackage{graphicx,rotating,subfigure,amsmath,amsfonts,amssymb,delarray,color}
\usepackage{hyperref}
\usepackage{xcolor}
\usepackage{soul}
\usepackage{dsfont}
\usepackage[T1]{fontenc}
\usepackage{physics}
\usepackage{multirow}

\newcommand{\e}{\text{e}}
\newcommand{\im}{\text{i}}

\def\12{\frac{1}{2}}

\begin{document}
\title{Hidden zero modes and topology of multiband non-Hermitian systems}
\author{Kyle Monkman$^{1}$}
\author{Jesko Sirker$^{2}$}
\affiliation{$^{1}$Stewart Blusson Quantum Matter Institute, University of British Columbia, Vancouver, British Columbia V6T 1Z4, Canada}
\affiliation{$^{2}$Department of Physics and Astronomy and Manitoba
Quantum Institute, University of Manitoba, Winnipeg, Canada R3T 2N2}
\date{\today}
\begin{abstract}
In a finite one-dimensional non-Hermitian system, the number of zero modes does not necessarily reflect the topology of the system. This is known as the breakdown of the bulk-boundary correspondence and has led to misconceptions about the topological protection of edge modes in such systems. Here we show why this breakdown does occur and that it typically results in hidden zero modes, extremely long-lived zero energy excitations, which are only revealed when considering the singular value instead of the eigenvalue spectrum. We point out, furthermore, that in a finite multiband non-Hermitian system with Hamiltonian $H$, one needs to consider also the reflected Hamiltonian $\tilde H$, which is in general distinct from the adjoint $H^\dagger$, to properly relate the number of protected zeroes to the winding number of $H$. 
\end{abstract}
\maketitle

\section{Introduction}
In quantum physics, we usually consider observables which are Hermitian operators. These operators have a real eigenspectrum, guaranteeing that expectation values are real and time evolution unitary. However, it has been pointed out about 20 years ago that one can replace the condition of Hermiticity by the less stringent condition of space-time reflection (PT) symmetry and, if this symmetry is unbroken, still obtain a real spectrum \cite{Bender_2005}. Another motivation to study non-Hermitian Hamiltonians comes from Master equations for open quantum systems \cite{L1976,BreuerPetruccione}. If one ignores quantum jumps, which is a reasonable approximation in certain limits, one can rewrite such a Master equation as a non-Hermitian Hamiltonian \cite{RPBC2022,MingantiMiranowicz}. This approach has been used in recent times to analyze and interpret experimental results in optical and magnetic systems \cite{MiriAlu,SuEstrecho,YangWang,SlimWanjura}. 
Non-Hermitian systems do show a number of phenomena which are not present in the Hermitian case. Most intriguingly, their eigenspectrum is extremely sensitive to small perturbations \cite{BoettcherSilbermann}. The best known example is the non-Hermitian skin effect: changing the boundary conditions from periodic (PBC) to open (OBC) can lead to a localization of a macroscopic number of states at the boundaries \cite{OkumaKawabata}. Another well known phenomenon are exceptional points. At these points, two eigenvalues and also their corresponding eigenvectors coalesce \cite{AGU2021,BBK2021}. This is, of course, not possible in a Hermitian system where the eigenvectors always form an orthogonal system. 

From a theoretical perspective, an important step to understand these phenomena is to classify Gaussian non-Hermitian systems \cite{BL2002,KSUS2019}. Here, topology plays a crucial role because it is robust against small perturbations which break translational invariance. In one-dimensional Hermitian systems, topological order is only possible if the system possesses additional symmetries. In the case of non-spatial symmetries---such as time reversal, particle-hole, and chiral symmetry---this leads to the tenfold classification of symmetry-protected topological (SPT) order \cite{RyuSchnyderReview}. In contrast, one-dimensional non-Hermitian systems can have a non-trivial topology even without additional symmetries because the eigenspectrum is complex and the determinant of the Bloch Hamiltonian $\det h(k):\, [0,2\pi)\to\mathbb{C}$ as function of momentum $k$ can have a non-zero winding number in the complex plane. It has been shown that if such a winding around a reference point $E$ in the complex plane exists for a single-band model then there will be a skin effect for open boundaries \cite{OkumaKawabata}. It has therefore been suggested that the standard bulk topological invariants are not useful to define a bulk-boundary correspondence for non-Hermitian systems and that they have to be replaced by other invariants \cite{YaoWang,KunstEdvardsson} or by an invariant based on a modified Hermitian Bloch Hamiltonian \cite{KSUS2019}. However, zero energy edge modes in finite non-Hermitian systems are, in general, fragile to perturbations \cite{BoettcherSilbermann,NobuyukiSato}, implying that they often lack topological protection thus making a bulk-boundary correspondence based on such modes questionable. It is crucial to note that "topological features must be properties of a translationally not invariant Hamiltonian" \cite{RyuSchnyder}. This is discussed further in App.~\ref{AppA}. As an alternative, the singular value spectrum has been put forward \cite{HerviouBardarson,BrunelliNunnenkamp, Porras1, Porras2}. The singular values $s_i$ of a system with Hamiltonian $H$ are the square roots of the eigenvalues of $H^\dagger H$. For a Hermitian system, we therefore have that $s_i=|\lambda_i|$ where $|\lambda_i|$ are the eigenvalues of $H$. In this case, the standard bulk-boundary correspondence also applies to the singular value spectrum. However, for a non-Hermitian $H$ the eigenvalue and the singular value spectrum are different. While the former is unstable to small perturbations, the latter is stable and important properties of the singular value spectrum can be directly inferred from the topological winding number.

In this article we will show why this is the case. We show, in particular, that non-Hermitian systems quite generally have hidden zero modes if they have a non-trivial topology. These are topologically protected edge modes with exact zero eigenvalues for semi-infinite boundaries which are, however, not present in the eigenspectrum and do not converge to zero with system size for a finite system. They do, however, get mapped exponentially close to zero by a finite-system Hamiltonian and thus represent extremely long-lived states which are physically highly relevant and are only distinguishable experimentally from true eigenstates in the dynamics at a time scale which increases linearly with system size. We will show, furthermore, that in the multiband case one has to consider not only the Hamiltonian $H$ but also the reflected Hamiltonian $\tilde{H}$. We will put the relation between topology, zero modes for semi-infinite boundaries, and protected singular values for finite systems which converge to zero with increasing system size, on a firm footing by using theorems known from the study of Toeplitz operators. Using several examples, we will highlight that the eigenspectrum is, in general, insensitive to the topology of the system while the singular value spectrum can be directly related to the winding number. In App.~\ref{AppA} we will show, furthermore, that the bulk-boundary correspondence put forward here fully explains the topological protection of stable edge modes in models considered previously in the literature. On the other hand, the approaches considered in Refs.~\cite{YaoWang,KunstEdvardsson,KSUS2019} do predict zero modes in open systems, however, these modes generically are, as we will show, related to translational invariance, unstable to small perturbations, and thus not topologically protected. 

\section{Topology and winding numbers}
In a tight-binding approximation for a non-interacting system, electrons moving on a periodic lattice can be described by
\begin{equation}
    \hat H=\sum_k\Psi^\dagger_k h(k) \Psi_k 
\end{equation}
where $\Psi_k=(c_k^1,c_k^2,\cdots,c_k^N)$ for a unit cell with $N$ elements, and $h(k)$ is the $N\times N$ Bloch Hamiltonian. If $\det(h(k))$ has no zeroes in $k\in [0,2\pi)$---which is also sometimes denoted as $\hat H$ having a point gap at $E=0$---then we can define a winding number w.r.t.~$E=0$ by
\begin{equation}
\label{winding}
    \mathcal{I}=
     \int_{0}^{2\pi}\!\! \frac{dk}{2\pi\im}\, \partial_k \ln \det h(k) 
    =  \int_{0}^{2\pi}\!\! \frac{dk}{2\pi\im} \sum_j \frac{\partial_k \lambda_j(k)}{\lambda_j(k)} \, .
\end{equation}
In the second line, we have assumed that $h(k)$ is diagonalizable with eigenvalues $\lambda_j(k)$, $j=1,\cdots,N$. Note that $\lambda_j(0)=\lambda_j(2\pi)$ is periodic. This implies, in particular, that if all eigenvalues are real then $\mathcal{I}=0$. A generic one-dimensional Hermitian system, i.e.~a system without additional symmetries, is thus always topologically trivial. The only way for a one-dimensional Hermitian system to have non-trivial topology is symmetry-protected topological (SPT) order. These symmetries lead to a block structure of $h(k)$ and topology can be defined based on the properties of these blocks. This leads to the tenfold classification scheme for non-spatial symmetries and to topological crystalline orders for spatial symmetries \cite{HughesProdan,FangGilbert,MonkmanSirker3,MonkmanSirker4}. In contrast, $\mathcal{I}\neq 0$ is a generic property of non-Hermitian systems which does not require the presence of additional symmetries. The bulk-boundary correspondence for SPT phases of Hermitian systems uses the bulk topological invariant to provide a lower bound on the number of protected edge modes in a system with boundaries. In a non-Hermitian system, on the other hand, which has a non-zero winding number around a reference energy $E$, we have the skin effect \cite{OkumaKawabata}. I.e., a macroscopic number of modes become localized in a system with boundaries but they are fragile with respect to small perturbations. 

\section{Topology in one-dimensional semi-infinite systems}
Here we want to elucidate in detail the proper bulk-boundary correspondence in the non-Hermitian case. First, we consider a semi-infinite system with a single boundary. We define Fourier coefficients $h_j=\frac{1}{2\pi}\int_{0}^{2\pi} h(k) e^{\im k j} dk$ which leads to the real-space matrix
\begin{eqnarray}
\label{finiteHn}
    H = \begin{pmatrix}
        h_0 & h_1 & h_2 & \dots \\
        h_{-1} & h_0 & h_1 & \dots \\
        h_{-2} & h_{-1} & h_0 & \dots \\
        \vdots & \vdots & \vdots & \ddots
    \end{pmatrix} 
\end{eqnarray}
where each of the $h_j$ matrices has size $N$ equal to that of the unit cell. $H$ thus has the form of a block Toeplitz operator. For such an operator, the index is defined as
\begin{equation}
\label{Gohberg}
    \text{ind}(H) = D(\text{ker}(H))-D(\text{ker}(H^\dagger)) 
\end{equation}
with $\text{ker}(H^\dagger)=\text{ker}(H^T)$ and $D$ denotes the dimension. Here $\text{ker}(H)=\{v|Hv=0\}$ is the kernel of $H$. For a Hermitian system, we always have $\text{ind}(H) =0$. The fact that the index can be non-zero tells us that the right and left eigenspectra of systems with semi-infinite boundary conditions are in general not the same. Furthermore, Gohberg's index theorem directly relates the winding number \eqref{winding} with the index \eqref{Gohberg} by \cite{BoettcherSilbermann}
\begin{equation}
\label{Gohberg2}
    \mathcal{I}=-\text{ind}(H) \, .
\end{equation}
Thus, the winding number $\mathcal{I}$ tells us directly about the difference in the number of right and left zero eigenvalues of $H$ but not their total numbers. This is similar to the Chern number in the Hermitian case which describes the difference between left and right propagating edge modes but does not fix their total numbers.

It is also very important to distinguish between the case where $h_j$ are $N\times N$ blocks with $N>1$ and the scalar case, $N=1$, where $h_j$ are just complex numbers. In the latter case, Coburn's lemma states that either $D(\text{ker}(H))=0$ or $D(\text{ker}(H^\dagger))=0$. I.e., in this case the sign of a non-zero winding number $\mathcal{I}$ tells us immediately whether $H$ has only right or only left zeroes and $|\mathcal{I}|$ gives the total number of such zeroes. 

\noindent \textit{Example 1:} 
To illustrate these points, we consider an example where $h(k)=1+2 \e^{-\im k}$ which means that $\mathcal{I}=-\text{ind}(H)=-1$ and the Fourier coefficients are $h_0=1$ and $h_1=2$ with all the other ones being equal to zero. Coburn's lemma then implies that $D(\text{ker}(H))=1$ and $D(\text{ker}(H^\dagger))=0$, i.e., $H$ has exactly one right zero mode. In this simple case, we can explicitly construct the right zero mode by considering $Hv=0$ with $v=\begin{pmatrix} v_1 & v_2 & \dots \end{pmatrix}^T$ and $v_j \in \mathbb{C}$. This leads to the recurrence relation for the vector coefficients $v_j+2v_{j+1}=0$. A normalized solution is given by $v_j=(-1)^{j-1}\sqrt{3}/2^j$ which means that the zero mode is exponentially localized. 

\section{Finite systems and hidden zero modes}
In a finite Hermitian system with non-trivial SPT order and system size $L$, zero modes can typically be easily identified as eigenvalues which scale as $|\lambda|\sim\exp(-L)$. This, however, is in general not the case in a non-Hermitian system. Here, zero modes can be hidden. This can be understood by considering the example discussed above. If we consider $H_L v=0$ for a finite matrix $H_L$ and vector $v=\begin{pmatrix} v_1 & v_2 & \dots & v_L\end{pmatrix}^T$ then the recurrence relation above is supplemented by the boundary condition $v_{L+1}=0$. In this case, the only solution is $v\equiv 0$ which is not a proper solution. The protected zero mode, which does exist in the thermodynamic limit, cannot be found by considering the eigenspectrum of a finite system. The only eigenstate in the finite case is $v=\begin{pmatrix} 1 & 0 & \dots  0 \end{pmatrix}^T$ with eigenvalue $E=1$.

More generally, we call a vector $v$ a hidden zero mode in a finite system with Hamiltonian $H_L$ if $H_L v\neq E v$ but
\begin{equation}
\label{limit}
    \lim_{L \to \infty} ||H_L v ||=0 \, .
\end{equation}
In the example, the vector $v=\begin{pmatrix} v_1 & v_2 & \dots & v_L\end{pmatrix}^T$ with $v_j=(-1)^{j-1}\sqrt{3}/2^j$ is a hidden zero mode with $\lim_{L \to \infty} ||H_L v ||=\lim_{L \to \infty}|v_L|=0$. 

Another issue which needs to be addressed is that a finite system has two boundaries. To study the properties of the second boundary we also need the real-space matrix corresponding to $h(-k)$ which is given by $\tilde{H}=H(h_j\to h_{-j})$ and describes the reflected Hamiltonian. In the scalar case $\tilde H=H^T$, however, this is no longer true in the block case. If the winding number of $H$ is $\mathcal{I}$ then the winding number of $\tilde H$ is $\tilde{\mathcal{I}}=-\mathcal{I}$.

\begin{figure}
    \centering
    \includegraphics[width=0.99\linewidth]{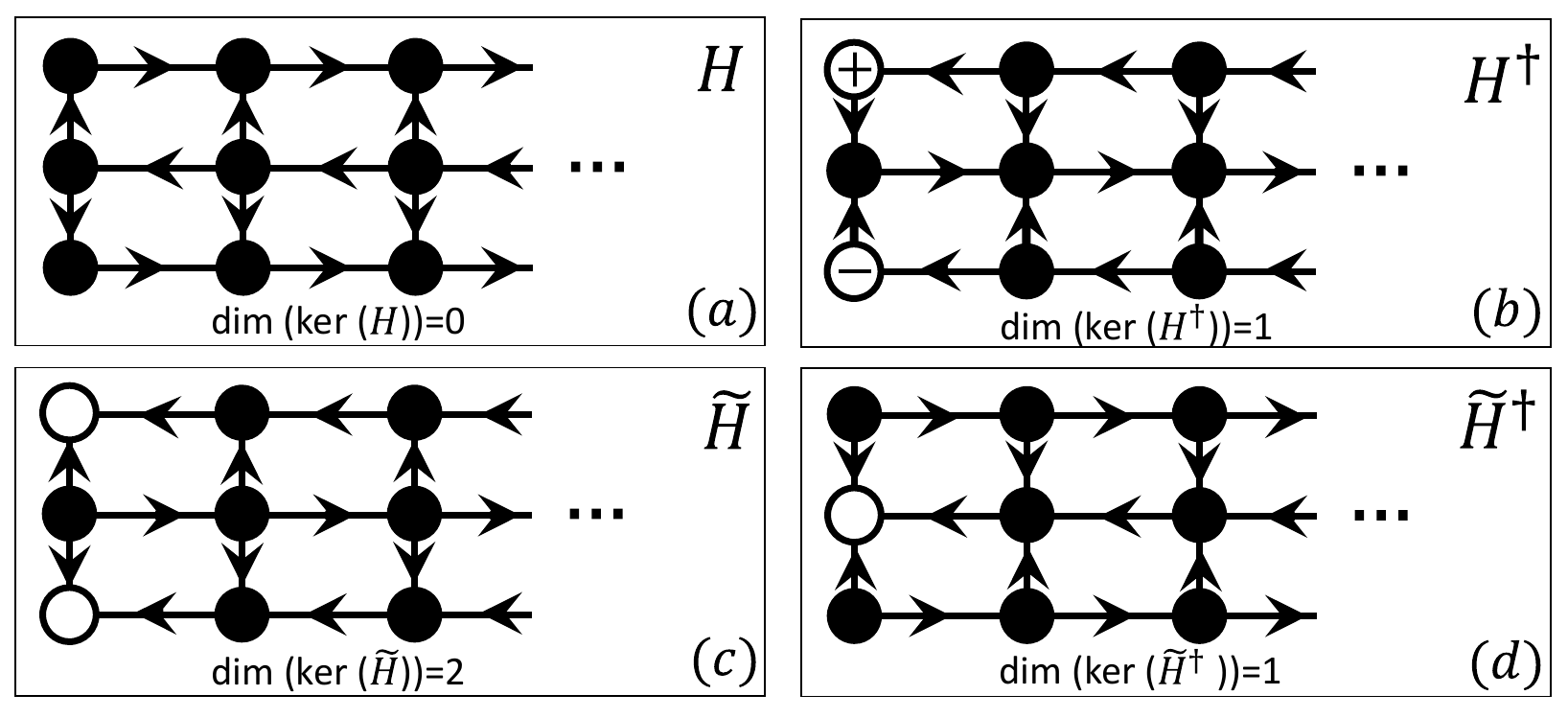}
    \caption{(a) The Hamiltonian $H$ in Eq.~\eqref{example2} and (b) its adjoint $H^\dagger$; (c) its reflection $\tilde H$ and (d) its adjoint $\tilde H^\dagger$. The arrows indicate unidirectional hopping and the open circles the localized edge modes.}
    \label{Fig1}
\end{figure}

\noindent \textit{Example 2:}
 This is best illustrated by another example. Consider the Hamiltonian in $k$-space given by
\begin{equation}
\label{example2}
    h(k) = \begin{pmatrix}
        \e^{ik} & 1 & 0 \\ 0 & \e^{-ik} & 0 \\ 0 & 1 & \e^{ik}
    \end{pmatrix} 
\end{equation}
with Fourier coefficients
\begin{equation}
\label{example2_2}
    h_0 = \begin{pmatrix}
        0 & 1 & 0 \\ 0 & 0 & 0 \\ 0 & 1 & 0
    \end{pmatrix} ,  
    h_{-1} = \begin{pmatrix}
        1 & 0 & 0 \\ 0 & 0 & 0 \\ 0 & 0 & 1
    \end{pmatrix} ,  h_1 = \begin{pmatrix}
        0 & 0 & 0 \\ 0 & 1 & 0 \\ 0 & 0 & 0
    \end{pmatrix} 
\end{equation} 
and $h_j=0$ otherwise. In this case with unidirectional hopping one obtains directly from $H$ that $D(\text{ker}(H))=0$ and $D(\text{ker}(\tilde{H}))=2$ with the two zero modes in the latter case being $\begin{pmatrix} 1& 0 &\dots &0 \end{pmatrix}^T$ and $\begin{pmatrix} 0& 0 & 1 &\dots &0 \end{pmatrix}^T$. Similarly, $D(\text{ker}(H^\dag))=1$ with zero mode $\begin{pmatrix} 1& 0 &-1 &\dots &0 \end{pmatrix}^T$ and $D(\text{ker}(\tilde{H}^\dag))=1$ with zero mode $\begin{pmatrix} 0& 1 &\dots &0 \end{pmatrix}^T$. This is illustrated in Fig.~\ref{Fig1}. This simple example shows that the zero modes of $H$ and $\tilde H$ are very different and that, in the non-scalar case, $\tilde H\neq H^T$. For the winding numbers it follows from the index theorem (\ref{Gohberg}, \ref{Gohberg2}) that $\mathcal{I}=D(\text{ker}(H^\dag)) - D(\text{ker}(H))=1$ and $\tilde{\mathcal{I}}=D(\text{ker}(\tilde{H}^\dag)) - D(\text{ker}(\tilde{H}))= -1$.
If we consider this Hamiltonian on a finite line then we find two exact zero modes which are localized at the right boundary. However, adding a small hopping in the opposite direction, one zero mode can be removed and the other one becomes a hidden zero mode.

\section{Singular values and hidden zero modes}
The instability of the eigenspectrum in the non-Hermitian case, and the issues of the two distinct boundaries and the hidden zero modes leads to the question how a bulk-boundary correspondence can be properly formulated. Here, the general theory of truncated Toeplitz operators provides an answer \cite{BoettcherSilbermann,FlynnViola,FlynnViola2}. While relatively little is known about the eigenspectrum of such operators, in particular in the non-scalar case, the $K$-splitting theorem \cite{Boettcher} directly relates the index of the operator, and thus also its winding number, to the spectrum of singular values. Assume we have a Hamiltonian $h(k)$ where $\det(h(k))$ has no zeroes in $k\in [0,2\pi)$. Then we define
\begin{equation}
    \label{k-splitting}
    K=D(\mbox{ker}(H)) + D(\mbox{ker}(\tilde H)) \, ,
\end{equation}
and the singular value spectrum $s_n(H_L)$ of a finite chain of length $L$ has the $K$-splitting property that
\begin{equation}
    \label{k-splitting2}
    \lim_{L\to\infty} s_n(H_L) = \left\{\begin{array}{rr} 0, & \quad\mbox 1\leq n\leq K\\ >0, & \quad n > K \, . \end{array}\right.
\end{equation}
I.e., there are exactly $K$ singular values which go to zero in the thermodynamic limit. Now we want to relate this property to the winding number \eqref{winding}. Let us start with the scalar case $N=1$. In this case, $\tilde H = H^T$ and $D(\mbox{ker}(\tilde H))=D(\mbox{ker}(H^\dagger))$. In addition, we also know from Coburn's lemma that either $D(\mbox{ker}(H))=0$ or $D(\mbox{ker}(H^\dagger))=0$. We can therefore also write $K=|D(\mbox{ker}(H))-D(\mbox{ker}(H^\dagger))|=|\mathcal{I}|$. I.e., in the scalar case there are exactly $|\mathcal{I}|$-many protected singular values. 

In the block case, $N>1$, both $D(\mbox{ker}(H))$ and $D(\mbox{ker}(H^\dagger))$ can be non-zero and the winding number $\mathcal{I}$ only fixes the difference. We know that $\tilde{\mathcal{I}}=-\mathcal{I}$ which implies that we can also write $K=D(\mbox{ker}(H^\dagger)) + D(\mbox{ker}(\tilde H^\dagger))$ and therefore
\begin{eqnarray}
    \label{k-splitting3}
    2K &=& D(\mbox{ker}(H)) + D(\mbox{ker}(\tilde H)) + D(\mbox{ker}(H^\dagger))  \nonumber \\
    &+& D(\mbox{ker}(\tilde H^\dagger)) \geq  |D(\mbox{ker}(H))-D(\mbox{ker}(H^\dagger))| \nonumber \\
    &+& |D(\mbox{ker}(\tilde H))-D(\mbox{ker}(\tilde H^\dagger))| = 2|\mathcal{I}| \, .
\end{eqnarray}
I.e., in the block case we only know that there are at least $|\mathcal{I}|\leq K$ singular values which will go to zero. Note that this is similar to the case of a Hermitian Chern insulator where the Chern number also only fixes the difference between left and right propagating modes but not their total number. The Chern number therefore also only provides a lower bound on the total number of edge modes.

Next, we want to connect the number of singular values $K$ going to zero with the total number of exact and hidden zero modes in the eigenspectrum. Consider the singular value decomposition $H_L=USV^\dagger$ with $U,V$ unitary and $S$ diagonal and positive. It follows that $H^\dagger_L H_L= VS^2V^\dagger$. Let us denote the column vectors of $V$ by $v_n$, those of $U$ by $u_n$ and the singular values on the diagonal of $S$ by $s_n$. We then obtain
\begin{equation}
    \label{hiddenzero}
    H_L^\dagger H_L v_n = s_n^2 v_n \quad \Rightarrow \quad H_L v_n = s_n u_n \, .
    \end{equation}
Now consider, in particular, one of the $K$ singular values with $\lim_{L\to\infty}s_n=0$. In this case we have
\begin{equation}
    \label{hiddenzero2}
    \lim_{L\to\infty} || H_L v_n || = \lim_{L\to\infty} ||s_n u_n|| = \lim_{L\to\infty}s_n=0 
\end{equation}
where we have used that $U$ is unitary. We thus find one of the main results of this letter: Every singular value $s_n$ which vanishes in the thermodynamic limit is directly connected to a (hidden) zero mode $v_n$ of the Hamiltonian $H_L$. The $K$-splitting theorem then connects this to the winding number. For a system with winding number $\mathcal{I}$ there are at least $|\mathcal{I}|$ exact or hidden zero modes. In the scalar case $N=1$, there are exactly $|\mathcal{I}|$ many which are protected by the winding number.

\begin{figure}[!t]
    \centering
    \includegraphics[width=0.99\columnwidth]{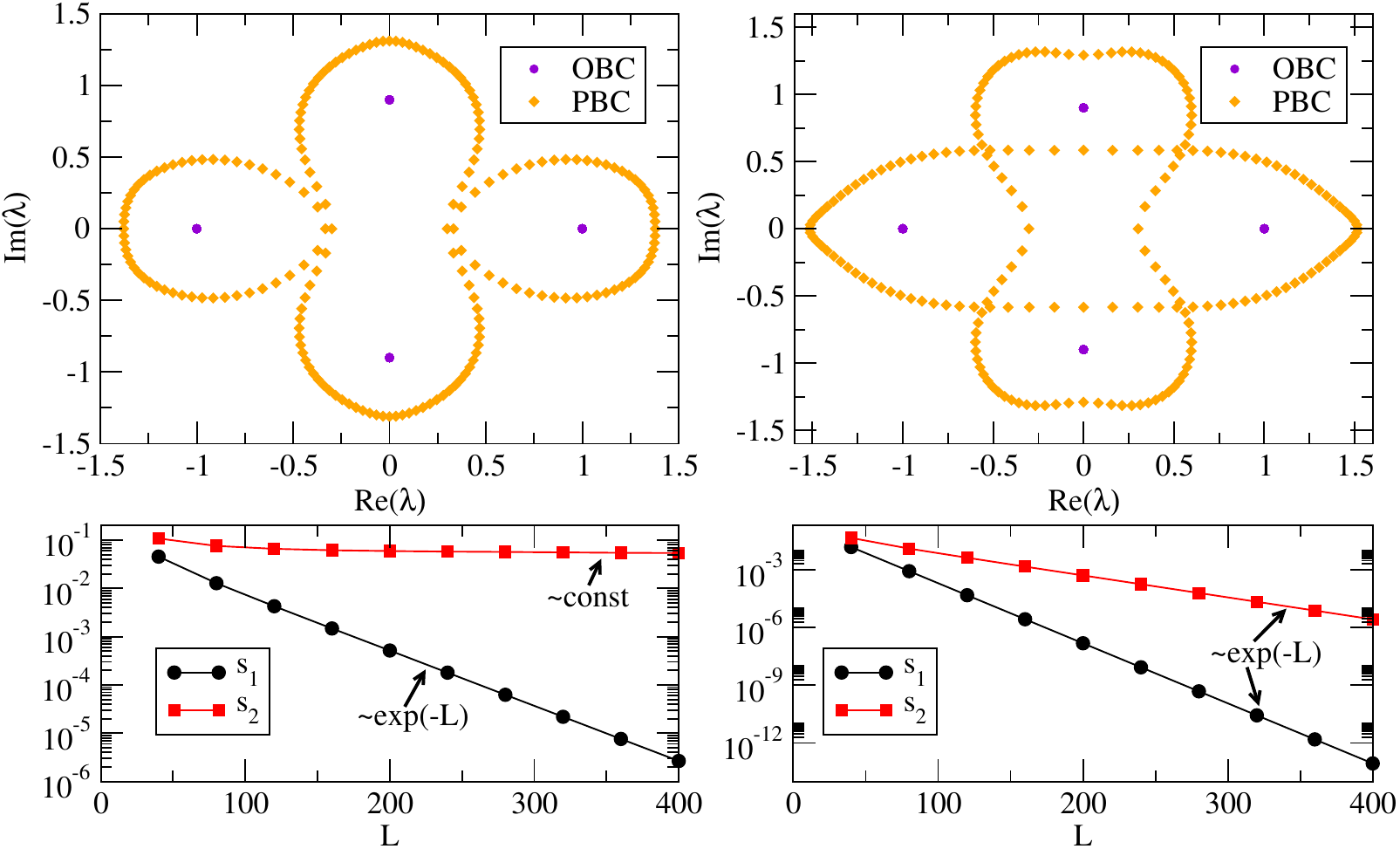}
    \caption{Left column: Case $\alpha=1$, $x=0.8$, $y=1$ with total winding $\mathcal{I}=1$. Right column: Case $\alpha=1$, $x=1.2$, $y=1$ with $\mathcal{I}=2$. The top row shows the PBC and the OBC spectra for $L=200$, the bottom row the smallest two singular values as a function of $L$. The OBC eigenspectrum is the same in both cases and insensitive to the change of topology.}
    \label{Fig_example_chiral1}
\end{figure}
\begin{figure}[!t]
    \centering
    \includegraphics[width=0.99\columnwidth]{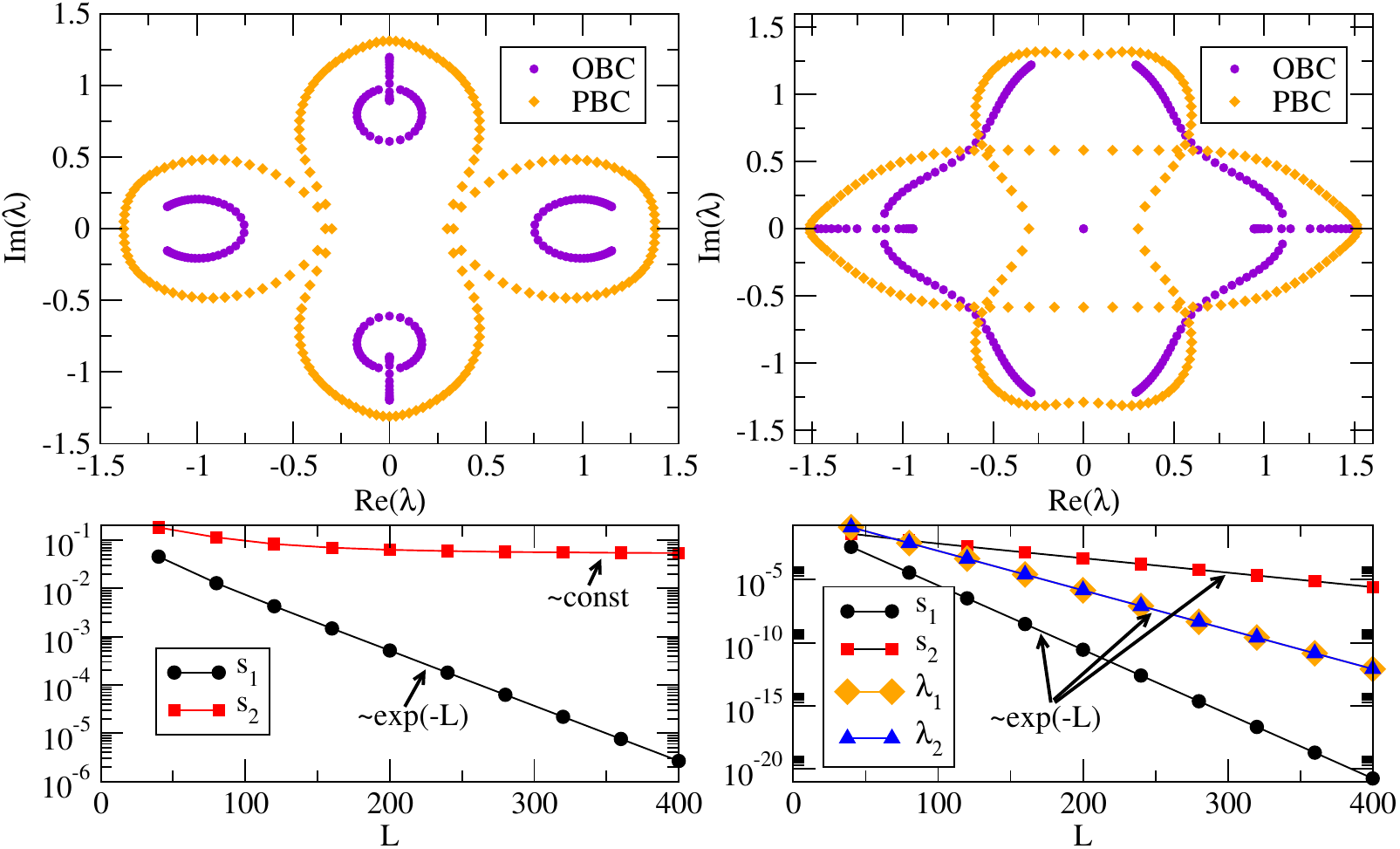}   
    \caption{Left column: Case $\alpha=-1$, $x=0.8$, $y=1$ with $\mathcal{I}=1$. Right column: Case $\alpha=-1$, $x=1.2$, $y=1$ with $\mathcal{I}=0$. The top row shows the PBC and the OBC spectra for $L=200$, the bottom row the smallest two singular values (and smallest two eigenvalues in the right panel). The zero mode in the left example is hidden while there are two eigenvalues going to zero in the right example.}
    \label{Fig_example_chiral2}
\end{figure}

\section{Hidden zero modes in a sub-lattice symmetric multiband model}
The following example shows several of these phenomena. Consider the Bloch Hamiltonian
\begin{equation}
    h(k)=\left(\begin{array}{cccc}
0 & 0 & 0.9 & \e^{ik} \\
0 & 0 & 1 & 1 \\
-0.9 & \e^{\alpha \im k} & 0 & 0 \\
x & y & 0 & 0
    \end{array}\right)   
\end{equation}
with $\alpha=\pm 1$ and real parameters $x>0,y>0$. The Hamiltonian has a sub-lattice symmetry with an upper right $2\times 2$ block $h_1(k)$ and a lower left $2\times 2$ block $h_2(k)$. We can define winding numbers for the two blocks which are, according to Eq.~\eqref{winding}, determined by their determinants with $\det h_1(k)=0.9-\e^{\im k}$ and $\det h_2(k)=-0.9y-x\e^{\alpha \im k}$.  The winding of the upper block is therefore $\mathcal{I}_1=1$. The lower block has winding $\mathcal{I}_2=\alpha$ if $x>0.9 y$ and no winding, $\mathcal{I}_2=0$, otherwise.

Let us first consider the case $\alpha=1$. In this case we can have a total winding $\mathcal{I}=\mathcal{I}_1+\mathcal{I}_2=2$ if $x>0.9y$ and $\mathcal{I}=1$ otherwise. In the former case, there are two singular values for the open system which are non-degenerate but both approach zero for system size $L\to \infty$. For the $\mathcal{I}=1$ case, on the other hand, there is only one singular value approaching zero for $L\to\infty$. The singular values thus clearly distinguish between these topologically distinct cases, see Fig.~\ref{Fig_example_chiral1}. In both cases there is a skin effect and all eigenstates are localized at the left boundary.  What makes this model a very illustrative example is that we can calculate the characteristic polynomial for the open system and $L/4\in\mathbb{N}$ unit cells exactly leading to
\begin{equation}
    [(\lambda-\sqrt{y})(\lambda+\sqrt{y})(\lambda-0.9\im)(\lambda+0.9\im)]^{L/4}=0 \, .
\end{equation}
This means that the eigenvalues are $\pm\sqrt{y}$ and $\pm 0.9\im$, each with a multiplicity of $L/4$ and {\it completely independent of $x$}. We can, in particular, switch between the cases $\mathcal{I}=2$ and $\mathcal{I}=1$ without changing the eigenvalue spectrum at all. It is completely insensitive to the change of topology. The zero modes in the eigenspectrum are only present in the semi-infinite system. For a finite system, these hidden modes can only be seen by considering the singular value spectrum as illustrated in Fig.~\ref{Fig_example_chiral1}. Note also, that in both cases there is no line gap. 

The case $\alpha=-1$ further exemplifies some of the findings in this work. Now $\mathcal{I}_2=-1$ if $x>0.9 y$ and $\mathcal{I}_2=0$ otherwise meaning that the total winding $\mathcal{I}$ is either zero or $1$. Naively, we might expect that for $\mathcal{I}=0$ the topology is trivial and there is no skin effect. This, however, is incorrect. With sub-lattice symmetry being present, there are two topological invariants which can be either considered to be the windings of the blocks $\mathcal{I}_{1,2}$ or the sum {\it and} the difference of these two invariants \cite{KSUS2019,YJLLC2018,JYC2018}. The number of singular values $K$ is bounded from below by $K\geq |\mathcal{I}_1|+|\mathcal{I}_2|$. We thus expect to have either two or just one protected singular value consistent with the numerical results shown in Fig.~\ref{Fig_example_chiral2}. However, the single zero mode for the $\mathcal{I}_1=1$, $\mathcal{I}_2=0$ case is hidden while two eigenvalues which go to zero are present for the $\mathcal{I}_1=1$, $\mathcal{I}_2=-1$ case. The latter can be explained by noting that the system for $\mathcal{I}_1=-\mathcal{I}_2$ can be adiabatically deformed to a Hermitian one. The two zero eigenmodes then follow from the standard bulk-boundary correspondence.

\section{Conclusions}
To conclude, we have shown why hidden zero modes appear in finite non-Hermitian systems and how a rigorous bulk-boundary correspondence can be established---based on well-known results for truncated Toeplitz operators---relating the winding number with the singular value spectrum. We also pointed out that zero eigenmodes in finite non-Hermitian systems are often not topologically protected, an issue which we discuss further in App.~\ref{AppA}, and that there are important differences between the scalar case and multiband models. In experiments, hidden zero modes reveal themselves as extremely long-lived states which can be detected, for example, by preparing the system in this state and let it time evolve \cite{NaghilooAbbasi}, see also App.~\ref{AppB}.
%which also contains Ref. \cite{MonkmanSirker2}.

Note added: After conclusion of this work, we became aware of Ref.~\cite{UghrelidzeViola} where the importance of the reflected Hamiltonian $\tilde{H}$ in the non-scalar case was pointed out as well.

Acknowledgements: The authors acknowledge support by NSERC via the Discovery grants program and by the DFG via the Research Unit FOR 2316.

\appendix 
\section{Appendix A: Other bulk-boundary correspondences}
\label{AppA}
The purpose of this appendix is to consider other examples which have been studied in the literature and which have lead to some misconceptions about what is and what is not topologically protected in a non-Hermitian system. 

\subsection{Non-Hermitian SSH model}
In Refs.~\cite{YaoWang,KunstEdvardsson} a non-Hermitian Su-Schrieffer-Heeger (SSH) model was considered which has the following $k$-space Hamiltonian
\begin{equation}
\label{H_PRL}
    h(k)=\left(\begin{array}{cc} 
    0 & t_1+\frac{\gamma}{2}+\exp(-\im k) \\
    t_1-\frac{\gamma}{2}+\exp(\im k) & 0  
    \end{array} \right)
\end{equation}
where $t_1$ and $\gamma$ are real parameters. This model has sub-lattice symmetry $\sigma_zh(k)\sigma_z=-h(k)$. We note that this is different from chiral symmetry $\sigma_zh(k)\sigma_z=-h^\dagger(k)$ which this model does not possess. The Fourier coefficients of this model are given by
\begin{equation}
    \label{F_PRL}
   h_0=\left(\begin{array}{cc} 
    0 & t_1+\frac{\gamma}{2} \\
    t_1-\frac{\gamma}{2} & 0  
    \end{array} \right)\! , 
    h_1=\left(\begin{array}{cc} 
    0 & 1 \\
    0 & 0  
    \end{array} \right)\! ,  
    h_{-1}=\left(\begin{array}{cc} 
    0 & 0 \\
    1 & 0  
    \end{array} \right)\! , 
\end{equation}
with all other coefficients being equal to zero. We have two topological invariants: the winding numbers of the upper right element $\mathcal{I}_1$ and of the lower left element $\mathcal{I}_2$. We see immediately that $\mathcal{I}_1=-1$ if $-1<t_1+\gamma/2<1$ and zero otherwise. Similarly, we have $\mathcal{I}_2=+1$ if $-1<t_1-\gamma/2<1$ and zero otherwise. The number of protected singular values is given by $K\geq |\mathcal{I}_1|+|\mathcal{I}_2|$ and we have three distinct regions with $|\mathcal{I}_1|+|\mathcal{I}_2|=2,1,0$.

\begin{figure}[!t]
    \centering
    \includegraphics[width=0.99\columnwidth]{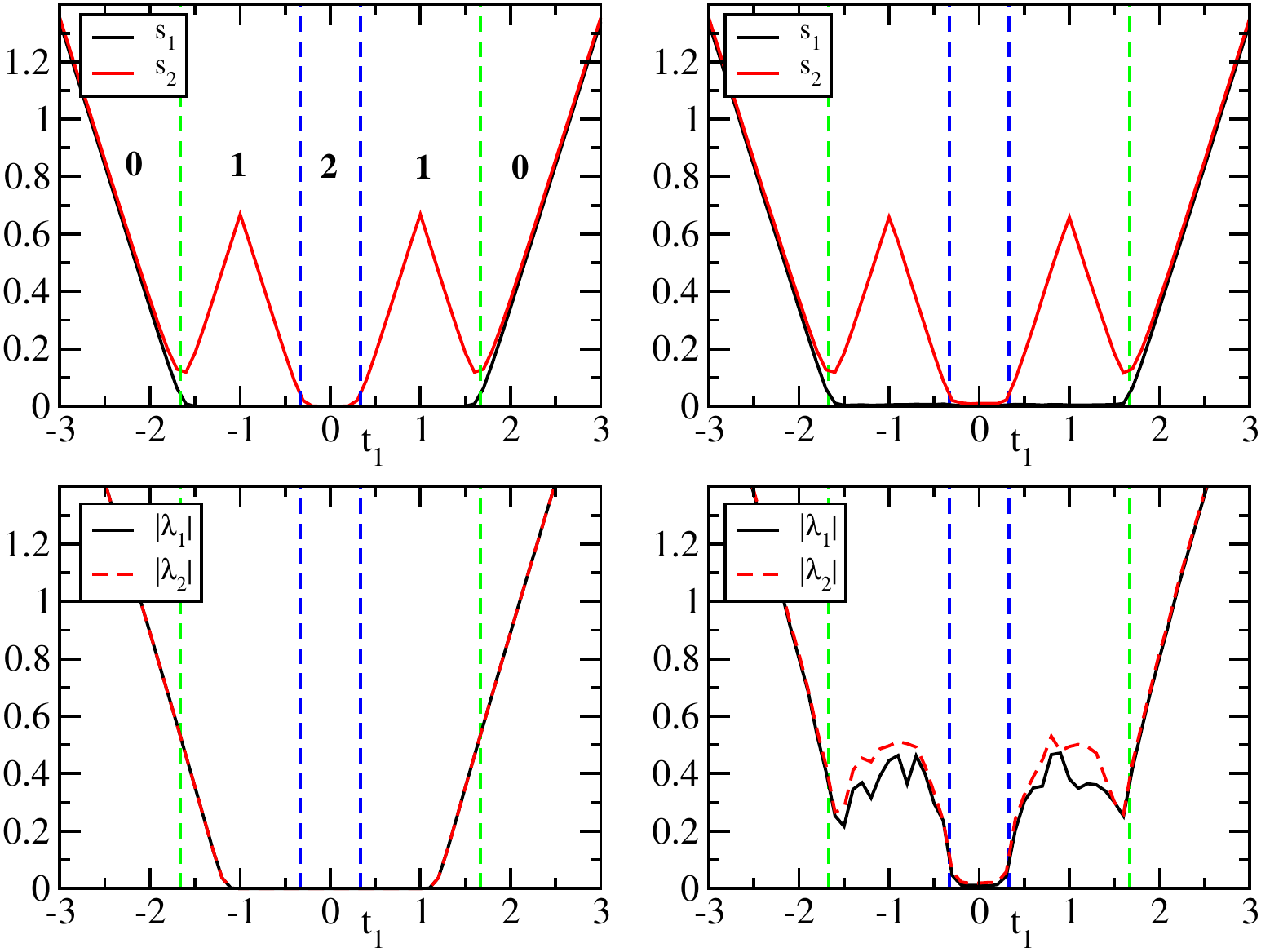}   
    \caption{Example from Fig.~2 in Ref.~\cite{YaoWang} for $40$ unit cells. Top row: The two smallest singular values, bottom row: The two eigenvalues of smallest magnitude. The left column shows the unperturbed system, the right column the perturbed system where a random matrix $M$ with elements $|m_{ij}|<10^{-2}$ has been added. At each point $t_1$, an average has been taken over ten such random matrices. The singular values are completely stable while the eigenvalues are only stable in $-1/3<t_1<1/3$.}
    \label{Fig_PRL}
\end{figure}

Let us now consider the specific example which has led to some confusion in Ref.~\cite{YaoWang}. The example studied in Ref.~\cite{KunstEdvardsson} is similar but uses a different $\gamma$ parameter which leads to a less complex phase diagram. For this reason we concentrate on Fig.~2 in Ref.~\cite{YaoWang} where $\gamma=4/3$ and $t_1$ is used as a parameter. In this specific case, $\mathcal{I}_1=-1$ if $-5/3<t_1<1/3$ and $\mathcal{I}_2=+1$ if $-1/3<t_1<5/3$. We thus expect that there is one protected singular value which goes to zero in the thermodynamic limit for a finite system if $1/3<|t_1|<5/3$ and two such protected singular values if $-1/3<t_1<1/3$. Outside of these regions, the model is topologically trivial and there are no protected modes. As shown in Fig.~\ref{Fig_PRL} for a system with $40$ unit cells, these results are in complete agreement with numerical calculations of the singular values. 

What has lead to some confusion is that there are two eigenvalues whose magnitude is going to zero with system size in a range which does not agree with the phase boundaries coming from the winding numbers. This has lead to the claim that one needs to abandon the notion of topology based on invariants for the Bloch Hamiltonian $h(k)$ \cite{YaoWang,KunstEdvardsson,KSUS2019}. In Ref.~\cite{YaoWang}, in particular, a 'non-Bloch topological invariant' is constructed which is meant to explain the topology of the system and the related topologically protected zero eigenmodes. This notion, however, is misguided because in part of the regime where the eigenvalues are zero there is no stability against small perturbations implying that there is no topological protection. The theory is based on translational invariance, however, "topological features must be properties of a translationally not invariant Hamiltonian" \cite{RyuSchnyder}. For the specific case considered, adding small random matrices to the Hamiltonian for a finite system clearly shows that the almost zero eigenvalues are only stable in $-1/3<t_1<1/3$, see Fig.~\ref{Fig_PRL} right column. Outside of this regime, there is no topological protection in contrast to what is implied by the non-Bloch invariant constructed in Ref.~\cite{YaoWang}. On the other hand, the singular values are completely stable against these perturbations demonstrating that they are indeed topologically protected. That there are topologically protected zero eigenmodes for $-1/3<t_1<1/3$ can be understood in the following manner: In this case we have $\mathcal{I}_1=-1$ and $\mathcal{I}_2=+1$ as for a chiral Hermitian system and, indeed, we can adiabatically deform the Hamiltonian \eqref{H_PRL} to a Hermitian one in this regime. This can be done, for example, by changing $\gamma/2\to -\gamma/2$ in the upper right block. Along this path, the gap never closes and the winding numbers thus remain the same as well. We conclude that the zero eigenmodes are only protected in the regime where the model is adiabatically connected to a topologically non-trivial Hermitian model. Outside of this regime, there is no protection of eigenmodes with zero energy.

This leaves us with the task to explain what happens in the regime $1/3<|t_1|<5/3$ where the winding numbers and singular values indicate that there is one protected mode for the semi-infinite chain. Based on the discussion in the main text, we expect again that for a finite system this will turn out to be a hidden zero mode. Because the Fourier coefficients \eqref{F_PRL} are so simple, we can explicitly demonstrate that this is indeed the case. Note that we have to consider again both $H$ and $\tilde{H}$ because the finite system has two boundaries. Let us start by considering $H$ for semi-infinite boundaries. We want to solve $Hv=0$. We then find for the odd vector coefficients
\begin{equation}
    \label{v1_PRL}
    v_1=0\quad\mbox{and}\quad v_{2j-1} + (t_1-\frac{\gamma}{2})v_{2j+1}=0 \quad\mbox{for}\; j=1,2,\cdots
\end{equation}
This implies that all odd coefficients are zero, $v_{2j-1}=0$ for $j=1,2,\cdots$. For the even vector coefficients we find the relation
\begin{equation}
    \label{v2_PRL}
    (t_1+\frac{\gamma}{2})v_{2j} + v_{2j+2}=0 \quad\mbox{for}\; j=1,2,\cdots
\end{equation}
We thus have $v_2$ as a free parameter and 
\begin{equation}
    \label{v22_PRL}
v_{2j}=\left(-t_1-\frac{\gamma}{2}\right)^{j-1}v_2\quad \mbox{for}\; j=2,3,\cdots. 
\end{equation}
However, for this to be a proper solution, we also have to demand that the solution is normalizable! We find $||v||^2=v_2^2\sum_{n=0}^\infty \left(t_1+\frac{\gamma}{2}\right)^{2n}$ which implies that we need $-1<t_1+\frac{\gamma}{2}<1$. This is exactly the regime where $\mathcal{I}_1=-1$. Thus we have proven that there is a zero mode in this regime which has the vector coefficients
\begin{equation}
    \label{v_PRL}
    v_{2j-1}=0,\; v_{2j}=\sqrt{1-(t_1+\gamma/2)^2}\left(-t_1-\frac{\gamma}{2}\right)^{j-1}
\end{equation}
for $j=1,2,\cdots$. In a finite system, we simply truncate the vector and we see that it becomes a hidden zero mode with $||Hv||\sim\exp(-L)$. 

For the second zero mode we have to consider the other boundary and thus $\tilde{H}$. Now we solve $\tilde{H}w=0$ and after a completely analogous calculation find that a normalizable solution exists if $-1<t_1-\frac{\gamma}{2}<1$ which is exactly the regime where $\mathcal{I}_2=+1$. The normalized solution in this regime reads
\begin{equation}
    \label{w_PRL}
    w_{2j}=0,\; w_{2j-1}=\sqrt{1-(t_1-\gamma/2)^2}\left(-t_1+\frac{\gamma}{2}\right)^{j-1}
\end{equation}
for $j=1,2,\cdots$. The truncated vector for a finite system is again a hidden zero mode with $||\tilde{H}w||\sim\exp(-L)$. 

We note that the same applies to the case $\gamma=3$ studied in Ref.~\cite{KunstEdvardsson}. The zero eigenmodes found in this case are unstable to perturbations as well. Instead, there is a stable hidden zero mode which exists in the regions $1/2<|t_1|<5/2$. While the polarization operator devised in Ref.~\cite{KunstEdvardsson} is quantized and detects the unstable zero modes by construction, it is in our view not correct to call this a bulk-boundary correspondence because the polarization operator requires the states of the open system as an input. The topological properties of the bulk Hamiltonian $h(k)$ do not enter at all. 

To conclude, we have shown that the winding numbers for $h(k)$ do predict the number of protected singular values which correspond to the number of topologically protected stable hidden and visible zero modes. The only regime where the zero modes for a finite system are protected and not hidden is the regime where the model can be adiabatically deformed to a Hermitian chiral model with $\mathcal{I}_1=-\mathcal{I}_2\neq 0$. In this case, the standard bulk-boundary correspondence applies. Outside of this regime, the eigenmodes with zero energy are accidental and unstable and not topologically protected in contradiction to Refs.~\cite{YaoWang,KunstEdvardsson}. Our theory of protected singular values and hidden zero modes, in contrast, provides the correct bulk-boundary correspondence also for this model.   

\subsection{Bloch Hamiltonian for open boundaries}
Another attempt at formulating a bulk-boundary correspondence for non-Hermitian systems has been to define a Bloch Hamiltonian for the bulk of an open system \cite{KSUS2019}. From a general perspective, this appears already questionable because for a non-Hermitian system there is no separation between bulk and boundary properties in the way we are used to for Hermitian systems. For example, the energies in a Hermitian system just aquire $1/L$ corrections when cutting a periodic chain and---except for possible edge states---changes to the eigenstates are limited to regions close to the boundary. In contrast, changing the boundary conditions from periodic to open typically completely changes all the eigenenergies and eigenstates of a non-Hermitian system. The corrections are not small in order $1/L$.

More specifically, it has been suggested that the spectrum and, in particular, the topological edge states in the model \eqref{H_PRL} for OBC can be understood based on the Bloch Hamiltonian
\begin{equation}
\label{H_PRX}
    h(k)=\left(\begin{array}{cc} 
    0 \!\!\!&\!\!\!\!\!\!\!\!\!\! \sqrt{t_1^2-\left(\frac{\gamma}{2}\right)^2}+\exp(-\im k) \\
    \sqrt{t_1^2-\left(\frac{\gamma}{2}\right)^2}+\exp(\im k) & 0  
    \end{array} \right).
\end{equation}
This Hamiltonian is chiral and Hermitian and thus will, according to the standard bulk-boundary correspondence \cite{MonkmanSirker2}, have two protected edge modes if $\mathcal{I}_1=-\mathcal{I}_2=-1$. This is the case if $|\sqrt{t_1^2-\left(\frac{\gamma}{2}\right)^2}|<1$. Clearly, this is very different from the conditions found for a single or two protected edge modes for the original model \eqref{H_PRL}. More specifically, the model \eqref{H_PRX} predicts that there are two protected edge modes if either (i) $|\gamma|/2<1$ and $-\sqrt{1+\left(\frac{\gamma}{2}\right)^2}<t_1<\sqrt{1+\left(\frac{\gamma}{2}\right)^2}$ or (ii) $|\gamma|/2>1$ and $\sqrt{-1+\left(\frac{\gamma}{2}\right)^2}<t_1<\sqrt{1+\left(\frac{\gamma}{2}\right)^2}$ or $-\sqrt{1+\left(\frac{\gamma}{2}\right)^2}<t_1<-\sqrt{-1+\left(\frac{\gamma}{2}\right)^2}$. For the example $\gamma=4/3$ considered in Fig.~\ref{Fig_PRL}, the chiral Hermitian Hamiltonian \eqref{H_PRX} thus has two protected edge modes if $-\sqrt{\frac{13}{9}}<t_1<\sqrt{\frac{13}{9}}$. This is indeed also the regime in which the eigenpectrum of the original model for OBC \eqref{F_PRL} has two zero eigenmodes. However, as we have aready shown in Fig.~\ref{Fig_PRL} these two zero modes are, in general, {\it not} topologically protected. It it thus not correct to relate the original non-Hermitian Bloch Hamiltonian to a chiral Hermitian one. This is only allowed if $\mathcal{I}_1=-\mathcal{I}_2$ in the original, non-Hermitian model. Topological protection in regimes where $\mathcal{I}_1\neq -\mathcal{I}_2$ is fundamentally different and cannot be captured by a Hermitian Bloch Hamiltonian. Instead, the topologically protected modes are hidden and only emerge as exact eigenstates in the thermodynamic limit.  

\section{Experimental detection of hidden zero modes}
\label{AppB}
In this appendix, we describe in more detail how hidden zero modes can be detected experimentally.

For a finite system $H_L$, a hidden zero mode $v$ is not an eigenmode, $H_L v\neq Ev$, but it does fulfill $||H_L v||\to 0$ for $L\to\infty$. Theoretically, we can find these zero modes either by using recurrence relations for the semi-infinite case when $v$ becomes an eigenmode with $Hv=0$ and then truncating the vector $v$ to length $L$ or by using a singular value decomposition. As we have shown in the main text, these two methods are equivalent. 

An experimental system will always be finite and a hidden zero mode will be close to but not quite an eigenstate. I.e., a hidden zero mode is a long-lived metastable state. Such states will show up in various response functions---in many-body physics we are used to excitations with a finite lifetime---but perhaps the easiest way to detect them is by a Loschmidt echo type experiment. In the simplest form of such an experiment, the system is prepared in the hidden zero mode $|\Psi\rangle$ and the generalized Loschmidt echo
\begin{equation}
\label{Loschmidt}
    \mathcal{L}(t)=\frac{\langle\Psi|\Psi(t)\rangle}{\sqrt{\langle\Psi(t)|\Psi(t)\rangle}}
\end{equation}
is obtained. $\mathcal{L}(t)$ is a measure for how long the time-evolved state $|\Psi(t)\rangle$ remains close to the initial state $|\Psi\rangle$. We call Eq.~\eqref{Loschmidt} a generalized Loschmidt echo because the time evolution is not unitary. Note that this implies, in particular, that the norm $\langle\Psi(t)|\Psi(t)\rangle$ is not conserved.

Another issue in an experiment is that the non-Hermitian description of an open quantum system is only effective and does ignore quantum jumps. I.e., for this description to work, we also need a postselection where the system is time-evolved many times and only those evolutions are taken into account where the system remains in the considered manifold. We note that exactly this type of experiment has already been performed for a qubit \cite{NaghilooAbbasi}. Here a qubit is prepared in an initial state and the probability monitored that, after postselection, the qubit remains in the initial state under time evolution with an effective non-Hermitian Hamiltonian. Such an experiment, generalized to multiple qubits, would be able to detect hidden zero modes. 

\begin{figure}[!t]
    \centering
    \includegraphics[width=0.99\columnwidth]{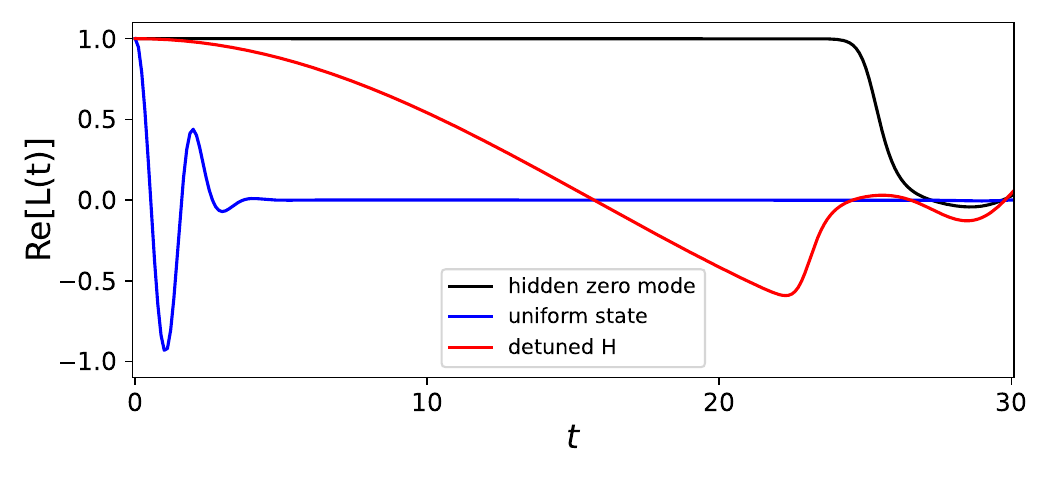} \\
    \includegraphics[width=0.99\columnwidth]{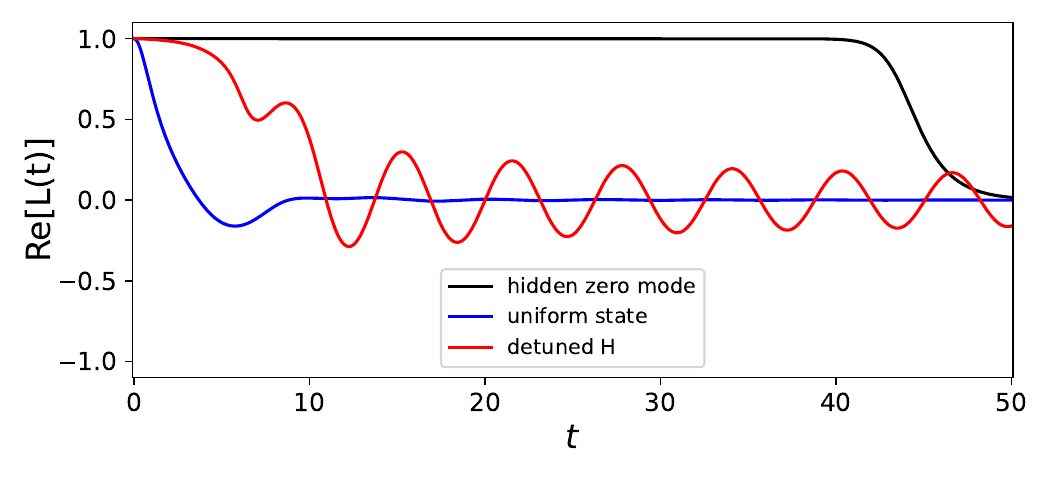}
    \caption{Upper panel: $\mathcal{L}(t)$ for the Hamiltonian in example 1 of the main paper. Lower panel: $\mathcal{L}(t)$ for the SSH chain, Eq.~\eqref{F_PRL}. In both cases the chain has $L=70$ sites.}
    \label{Fig_Loschmidt}
\end{figure}

To show the expected results, we calculate the Loschmidt echo \eqref{Loschmidt} for example 1 of the main paper as well as for the SSH model studied in the previous section. The results are shown in Fig.~\ref{Fig_Loschmidt}.
For example 1, the Hamiltonian has elements $h_0=1$, $h_1=2$ and the hidden zero mode is given by $v=\begin{pmatrix}\frac{\sqrt{3}}{2}, & - \frac{\sqrt{3}}{4}, &\dots, & \frac{\sqrt{3}}{2^{L-1}},& -\frac{\sqrt{3}}{2^L}\end{pmatrix}$. If we prepare the system in this state $v$, then $\mbox{Re}[\mathcal{L}(t)]\approx 1$ for times $gt<L$ where $g$ is some effective velocity which depends on the hopping amplitude. In contrast, if we prepare the system in the uniform state  $\tilde v=\begin{pmatrix}\frac{1}{\sqrt{L}}, & \frac{1}{\sqrt{L}}, &\dots, & \frac{1}{\sqrt{L}},& \frac{1}{\sqrt{L}}\end{pmatrix}$ then we see a damped oscillation and the Loschmidt echo goes to zero quickly. Also shown in Fig.~\ref{Fig_Loschmidt} is the case where the Hamiltonian is detuned by setting $h_1=2.2$ but keeping the initial state $v$. Then, there is some immediate decay clearly indicating that $v$ is not a hidden zero mode of the detuned Hamiltonian. 

For the SSH model we found that one of the hidden zero modes $v$ in the regime $-1<t_1+\frac{\gamma}{2}<1$ is given by Eq.~\eqref{v_PRL}. We prepare the system in this state for $t_1=-0.2$ and $\gamma=4/3$. Again, the Loschmidt echo shows that this mode is indeed a long-lived metastable state. We also show again $\mathcal{L}(t)$ for additional cases. One additional case is for the uniform state $\tilde v$ under the same Hamiltonian evolution. The other is for the hidden zero mode $v$ in the presence of a detuned Hamiltonian with the hopping amplitude changed to $t_1=-0.1$. In both cases, the Loschmidt echo quickly moves away from being equal to one.

We conclude that hidden zero modes are important features of non-Hermitian systems and are experimentally observable as long-lived states. They are distinguishable from true eigenstates because the Loschmidt echo does start to decay at a time scale which increases linearly with system size $L$ whereas $\mathcal{L}(t)=1$ for all times for a true eigenstate.


\begin{thebibliography}{36}%
\makeatletter
\providecommand \@ifxundefined [1]{%
 \@ifx{#1\undefined}
}%
\providecommand \@ifnum [1]{%
 \ifnum #1\expandafter \@firstoftwo
 \else \expandafter \@secondoftwo
 \fi
}%
\providecommand \@ifx [1]{%
 \ifx #1\expandafter \@firstoftwo
 \else \expandafter \@secondoftwo
 \fi
}%
\providecommand \natexlab [1]{#1}%
\providecommand \enquote  [1]{``#1''}%
\providecommand \bibnamefont  [1]{#1}%
\providecommand \bibfnamefont [1]{#1}%
\providecommand \citenamefont [1]{#1}%
\providecommand \href@noop [0]{\@secondoftwo}%
\providecommand \href [0]{\begingroup \@sanitize@url \@href}%
\providecommand \@href[1]{\@@startlink{#1}\@@href}%
\providecommand \@@href[1]{\endgroup#1\@@endlink}%
\providecommand \@sanitize@url [0]{\catcode `\\12\catcode `\$12\catcode
  `\&12\catcode `\#12\catcode `\^12\catcode `\_12\catcode `\%12\relax}%
\providecommand \@@startlink[1]{}%
\providecommand \@@endlink[0]{}%
\providecommand \url  [0]{\begingroup\@sanitize@url \@url }%
\providecommand \@url [1]{\endgroup\@href {#1}{\urlprefix }}%
\providecommand \urlprefix  [0]{URL }%
\providecommand \Eprint [0]{\href }%
\providecommand \doibase [0]{https://doi.org/}%
\providecommand \selectlanguage [0]{\@gobble}%
\providecommand \bibinfo  [0]{\@secondoftwo}%
\providecommand \bibfield  [0]{\@secondoftwo}%
\providecommand \translation [1]{[#1]}%
\providecommand \BibitemOpen [0]{}%
\providecommand \bibitemStop [0]{}%
\providecommand \bibitemNoStop [0]{.\EOS\space}%
\providecommand \EOS [0]{\spacefactor3000\relax}%
\providecommand \BibitemShut  [1]{\csname bibitem#1\endcsname}%
\let\auto@bib@innerbib\@empty
%</preamble>
\bibitem [{\citenamefont {Bender}(2005)}]{Bender_2005}%
  \BibitemOpen
  \bibfield  {author} {\bibinfo {author} {\bibfnamefont {C.~M.}\ \bibnamefont
  {Bender}},\ }\bibfield  {title} {\bibinfo {title} {Introduction to
  pt-symmetric quantum theory},\ }\href
  {https://doi.org/10.1080/00107500072632} {\bibfield  {journal} {\bibinfo
  {journal} {Contemp. Phys.}\ }\textbf {\bibinfo {volume} {46}},\
  \bibinfo {pages} {277} (\bibinfo {year} {2005})}\BibitemShut {NoStop}%
\bibitem [{\citenamefont {Lindblad}(1976)}]{L1976}%
  \BibitemOpen
  \bibfield  {author} {\bibinfo {author} {\bibfnamefont {G.}~\bibnamefont
  {Lindblad}},\ }\bibfield  {title} {\bibinfo {title} {On the generators of
  quantum dynamical semigroups},\ }\href
  {https://doi.org/https://doi.org/10.1007/BF01608499} {\bibfield  {journal}
  {\bibinfo  {journal} {Comm. Math. Phys.}\ }\textbf
  {\bibinfo {volume} {48}},\ \bibinfo {pages} {119} (\bibinfo {year}
  {1976})}\BibitemShut {NoStop}%
\bibitem [{\citenamefont {Breuer}\ and\ \citenamefont
  {Petruccione}(2002)}]{BreuerPetruccione}%
  \BibitemOpen
  \bibfield  {author} {\bibinfo {author} {\bibfnamefont {H.~P.}\ \bibnamefont
  {Breuer}}\ and\ \bibinfo {author} {\bibfnamefont {F.}~\bibnamefont
  {Petruccione}},\ }\href@noop {} {\emph {\bibinfo {title} {The Theory of Open
  Quantum Systems}}}\ (\bibinfo  {publisher} {Oxford University Press},\
  \bibinfo {address} {New York},\ \bibinfo {year} {2002})\BibitemShut {NoStop}%
\bibitem [{\citenamefont {Roccati}\ \emph {et~al.}(2022)\citenamefont
  {Roccati}, \citenamefont {Palma}, \citenamefont {Bagarello},\ and\
  \citenamefont {Ciccarello}}]{RPBC2022}%
  \BibitemOpen
  \bibfield  {author} {\bibinfo {author} {\bibfnamefont {F.}~\bibnamefont
  {Roccati}}, \bibinfo {author} {\bibfnamefont {G.~M.}\ \bibnamefont {Palma}},
  \bibinfo {author} {\bibfnamefont {F.}~\bibnamefont {Bagarello}},\ and\
  \bibinfo {author} {\bibfnamefont {F.}~\bibnamefont {Ciccarello}},\ }\bibfield
   {title} {\bibinfo {title} {{Non-Hermitian Physics and Master Equations}},\
  }\href {https://doi.org/10.1142/S1230161222500044} {\bibfield  {journal}
  {\bibinfo  {journal} {Open Systems \& Information Dynamics}\ }\textbf
  {\bibinfo {volume} {29}},\ \bibinfo {pages} {2250004} (\bibinfo {year}
  {2022})},\  \BibitemShut {NoStop}%
\bibitem [{\citenamefont {Minganti}\ \emph {et~al.}(2019)\citenamefont
  {Minganti}, \citenamefont {Miranowicz}, \citenamefont {Chhajlany},\ and\
  \citenamefont {Nori}}]{MingantiMiranowicz}%
  \BibitemOpen
  \bibfield  {author} {\bibinfo {author} {\bibfnamefont {F.}~\bibnamefont
  {Minganti}}, \bibinfo {author} {\bibfnamefont {A.}~\bibnamefont
  {Miranowicz}}, \bibinfo {author} {\bibfnamefont {R.~W.}\ \bibnamefont
  {Chhajlany}},\ and\ \bibinfo {author} {\bibfnamefont {F.}~\bibnamefont
  {Nori}},\ }\bibfield  {title} {\bibinfo {title} {Quantum exceptional points
  of non-hermitian hamiltonians and liouvillians: The effects of quantum
  jumps},\ }\href {https://doi.org/10.1103/PhysRevA.100.062131} {\bibfield
  {journal} {\bibinfo  {journal} {Phys. Rev. A}\ }\textbf {\bibinfo {volume}
  {100}},\ \bibinfo {pages} {062131} (\bibinfo {year} {2019})}\BibitemShut
  {NoStop}%
\bibitem [{\citenamefont {Miri}\ and\ \citenamefont {AlÃ¹}(2019)}]{MiriAlu}%
  \BibitemOpen
  \bibfield  {author} {\bibinfo {author} {\bibfnamefont {M.-A.}\ \bibnamefont
  {Miri}}\ and\ \bibinfo {author} {\bibfnamefont {A.}~\bibnamefont {Alu}},\
  }\bibfield  {title} {\bibinfo {title} {Exceptional points in optics and
  photonics},\ }\href {https://doi.org/10.1126/science.aar7709} {\bibfield
  {journal} {\bibinfo  {journal} {Science}\ }\textbf {\bibinfo {volume}
  {363}},\ \bibinfo {pages} {eaar7709} (\bibinfo {year} {2019})}\BibitemShut
  {NoStop}%
\bibitem [{\citenamefont {Su}\ \emph {et~al.}(2021)\citenamefont {Su},
  \citenamefont {Estrecho}, \citenamefont {BiegaÅska}, \citenamefont {Huang},
  \citenamefont {Wurdack}, \citenamefont {Pieczarka}, \citenamefont {Truscott},
  \citenamefont {Liew}, \citenamefont {Ostrovskaya},\ and\ \citenamefont
  {Xiong}}]{SuEstrecho}%
  \BibitemOpen
  \bibfield  {author} {\bibinfo {author} {\bibfnamefont {R.}~\bibnamefont
  {Su}}, \bibinfo {author} {\bibfnamefont {E.}~\bibnamefont {Estrecho}},
  \bibinfo {author} {\bibfnamefont {D.}~\bibnamefont {Biegaska}}, \bibinfo
  {author} {\bibfnamefont {Y.}~\bibnamefont {Huang}}, \bibinfo {author}
  {\bibfnamefont {M.}~\bibnamefont {Wurdack}}, \bibinfo {author} {\bibfnamefont
  {M.}~\bibnamefont {Pieczarka}}, \bibinfo {author} {\bibfnamefont {A.~G.}\
  \bibnamefont {Truscott}}, \bibinfo {author} {\bibfnamefont {T.~C.~H.}\
  \bibnamefont {Liew}}, \bibinfo {author} {\bibfnamefont {E.~A.}\ \bibnamefont
  {Ostrovskaya}},\ and\ \bibinfo {author} {\bibfnamefont {Q.}~\bibnamefont
  {Xiong}}, \ } \bibfield  {title} {\bibinfo {title} {Direct measurement of a
  non-hermitian topological invariant in a hybrid light-matter system},\ }\href
  {https://doi.org/10.1126/sciadv.abj8905} {\bibfield  {journal} {\bibinfo
  {journal} {Sci. Adv.}\ }\textbf {\bibinfo {volume} {7}},\ \bibinfo
  {pages} {eabj8905} (\bibinfo {year} {2021})},\  \BibitemShut
  {NoStop}%
\bibitem [{\citenamefont {Yang}\ \emph {et~al.}(2020)\citenamefont {Yang},
  \citenamefont {Wang}, \citenamefont {Rao}, \citenamefont {Gui}, \citenamefont
  {Yao}, \citenamefont {Lu},\ and\ \citenamefont {Hu}}]{YangWang}%
  \BibitemOpen
  \bibfield  {author} {\bibinfo {author} {\bibfnamefont {Y.}~\bibnamefont
  {Yang}}, \bibinfo {author} {\bibfnamefont {Y.-P.}\ \bibnamefont {Wang}},
  \bibinfo {author} {\bibfnamefont {J.~W.}\ \bibnamefont {Rao}}, \bibinfo
  {author} {\bibfnamefont {Y.~S.}\ \bibnamefont {Gui}}, \bibinfo {author}
  {\bibfnamefont {B.~M.}\ \bibnamefont {Yao}}, \bibinfo {author} {\bibfnamefont
  {W.}~\bibnamefont {Lu}},\ and\ \bibinfo {author} {\bibfnamefont {C.-M.}\
  \bibnamefont {Hu}},\ }\bibfield  {title} {\bibinfo {title} {Unconventional
  singularity in anti-parity-time symmetric cavity magnonics},\ }\href
  {https://doi.org/10.1103/PhysRevLett.125.147202} {\bibfield  {journal}
  {\bibinfo  {journal} {Phys. Rev. Lett.}\ }\textbf {\bibinfo {volume} {125}},\
  \bibinfo {pages} {147202} (\bibinfo {year} {2020})}\BibitemShut {NoStop}% 
 \bibitem{SlimWanjura}%
  \BibitemOpen
  \bibfield  {author} {\bibinfo {author} {\bibfnamefont {J.~J.}~\bibnamefont
  {Slim}}, \bibinfo {author} {\bibfnamefont {C.~C.}\ \bibnamefont {Wanjura}},
  \bibinfo {author} {\bibfnamefont {M.}\ \bibnamefont {Brunelli}},
  \bibinfo {author} {\bibfnamefont {J.}\ \bibnamefont {del Pino}},
  \bibinfo {author} {\bibfnamefont {A.}\ \bibnamefont {Nunnenkamp}}, \ and\ \bibinfo {author} {\bibfnamefont {E.}\
  \bibnamefont {Verhagen}},\ }\bibfield  {title} {\bibinfo {title} {Optomechanical realization of the bosonic Kitaev chain},\ }\href
  {https://doi.org/10.1038/s41586-024-07174-w} {\bibfield  {journal}
  {\bibinfo  {journal} {Nature}\ }\textbf {\bibinfo {volume} {627}},\
  \bibinfo {pages} {767} (\bibinfo {year} {2024})}\BibitemShut {NoStop}%
\bibitem [{\citenamefont {B{\"o}ttcher}\ and\ \citenamefont
  {Silbermann}(1999)}]{BoettcherSilbermann}%
  \BibitemOpen
  \bibfield  {author} {\bibinfo {author} {\bibfnamefont {A.}~\bibnamefont
  {B{\"o}ttcher}}\ and\ \bibinfo {author} {\bibfnamefont {B.}~\bibnamefont
  {Silbermann}},\ }\href@noop {} {\emph {\bibinfo {title} {Introduction to
  large truncated Toeplitz matrices}}}\ (\bibinfo  {publisher} {Springer (New
  York)},\ \bibinfo {year} {1999})\BibitemShut {NoStop}%
\bibitem [{\citenamefont {Okuma}\ \emph {et~al.}(2020)\citenamefont {Okuma},
  \citenamefont {Kawabata}, \citenamefont {Shiozaki},\ and\ \citenamefont
  {Sato}}]{OkumaKawabata}%
  \BibitemOpen
  \bibfield  {author} {\bibinfo {author} {\bibfnamefont {N.}~\bibnamefont
  {Okuma}}, \bibinfo {author} {\bibfnamefont {K.}~\bibnamefont {Kawabata}},
  \bibinfo {author} {\bibfnamefont {K.}~\bibnamefont {Shiozaki}},\ and\
  \bibinfo {author} {\bibfnamefont {M.}~\bibnamefont {Sato}},\ }\bibfield
  {title} {\bibinfo {title} {Topological origin of non-hermitian skin
  effects},\ }\href {https://doi.org/10.1103/PhysRevLett.124.086801} {\bibfield
   {journal} {\bibinfo  {journal} {Phys. Rev. Lett.}\ }\textbf {\bibinfo
  {volume} {124}},\ \bibinfo {pages} {086801} (\bibinfo {year}
  {2020})}\BibitemShut {NoStop}%
\bibitem [{\citenamefont {Ashida}\ \emph {et~al.}(2021)\citenamefont {Ashida},
  \citenamefont {Gong},\ and\ \citenamefont {Ueda}}]{AGU2021}%
  \BibitemOpen
  \bibfield  {author} {\bibinfo {author} {\bibfnamefont {Y.}~\bibnamefont
  {Ashida}}, \bibinfo {author} {\bibfnamefont {Z.}~\bibnamefont {Gong}},\ and\
  \bibinfo {author} {\bibfnamefont {M.}~\bibnamefont {Ueda}},\ }\bibfield
  {title} {\bibinfo {title} {{Non-Hermitian physics}},\ }\href
  {https://doi.org/10.1080/00018732.2021.1876991} {\bibfield  {journal}
  {\bibinfo  {journal} {Adv. Phys.}\ }\textbf {\bibinfo {volume} {69}},\
  \bibinfo {pages} {249} (\bibinfo {year} {2021})},\   \BibitemShut {NoStop}%
\bibitem [{\citenamefont {Bergholtz}\ \emph {et~al.}(2021)\citenamefont
  {Bergholtz}, \citenamefont {Budich},\ and\ \citenamefont {Kunst}}]{BBK2021}%
  \BibitemOpen
  \bibfield  {author} {\bibinfo {author} {\bibfnamefont {E.~J.}\ \bibnamefont
  {Bergholtz}}, \bibinfo {author} {\bibfnamefont {J.~C.}\ \bibnamefont
  {Budich}},\ and\ \bibinfo {author} {\bibfnamefont {F.~K.}\ \bibnamefont
  {Kunst}},\ }\bibfield  {title} {\bibinfo {title} {{Exceptional topology of
  non-Hermitian systems}},\ }\href
  {https://doi.org/10.1103/revmodphys.93.015005} {\bibfield  {journal}
  {\bibinfo  {journal} {Rev. Mod. Phys.}\ }\textbf {\bibinfo {volume} {93}},\
  \bibinfo {pages} {015005} (\bibinfo {year} {2021})},\ \Eprint
  {https://arxiv.org/abs/1912.10048} {arXiv:1912.10048 [cond-mat.mes-hall]}
  \BibitemShut {NoStop}%
\bibitem [{\citenamefont {Bernard}\ and\ \citenamefont
  {LeClair}(2002)}]{BL2002}%
  \BibitemOpen
  \bibfield  {author} {\bibinfo {author} {\bibfnamefont {D.}~\bibnamefont
  {Bernard}}\ and\ \bibinfo {author} {\bibfnamefont {A.}~\bibnamefont
  {LeClair}},\ }\bibinfo {title} {A classification of non-hermitian random
  matrices},\ in\ \href {https://doi.org/10.1007/978-94-010-0514-2_19} {\emph
  {\bibinfo {booktitle} {Statistical Field Theories}}}\ (\bibinfo  {publisher}
  {Springer Netherlands},\ \bibinfo {year} {2002})\ p.\ \bibinfo {pages}
  {207}\BibitemShut {NoStop}%
\bibitem [{\citenamefont {{Kawabata}}\ \emph {et~al.}(2019)\citenamefont
  {{Kawabata}}, \citenamefont {{Shiozaki}}, \citenamefont {{Ueda}},\ and\
  \citenamefont {{Sato}}}]{KSUS2019}%
  \BibitemOpen
  \bibfield  {author} {\bibinfo {author} {\bibfnamefont {K.}~\bibnamefont
  {{Kawabata}}}, \bibinfo {author} {\bibfnamefont {K.}~\bibnamefont
  {{Shiozaki}}}, \bibinfo {author} {\bibfnamefont {M.}~\bibnamefont {{Ueda}}},\
  and\ \bibinfo {author} {\bibfnamefont {M.}~\bibnamefont {{Sato}}},\
  }\bibfield  {title} {\bibinfo {title} {{Symmetry and topology in
  non-hermitian physics}},\ }\href {https://doi.org/10.1103/PhysRevX.9.041015}
  {\bibfield  {journal} {\bibinfo  {journal} {Physical Review X}\ }\textbf
  {\bibinfo {volume} {9}},\ \bibinfo {eid} {041015} (\bibinfo {year} {2019})}
   \BibitemShut {NoStop}%
\bibitem [{\citenamefont {Chiu}\ \emph {et~al.}(2016)\citenamefont {Chiu},
  \citenamefont {Teo}, \citenamefont {Schnyder},\ and\ \citenamefont
  {Ryu}}]{RyuSchnyderReview}%
  \BibitemOpen
  \bibfield  {author} {\bibinfo {author} {\bibfnamefont {C.-K.}\ \bibnamefont
  {Chiu}}, \bibinfo {author} {\bibfnamefont {J.~C.~Y.}\ \bibnamefont {Teo}},
  \bibinfo {author} {\bibfnamefont {A.~P.}\ \bibnamefont {Schnyder}},\ and\
  \bibinfo {author} {\bibfnamefont {S.}~\bibnamefont {Ryu}},\ }\bibfield
  {title} {\bibinfo {title} {Classification of topological quantum matter with
  symmetries},\ }\href {https://doi.org/10.1103/RevModPhys.88.035005}
  {\bibfield  {journal} {\bibinfo  {journal} {Rev. Mod. Phys.}\ }\textbf
  {\bibinfo {volume} {88}},\ \bibinfo {pages} {035005} (\bibinfo {year}
  {2016})}\BibitemShut {NoStop}%
\bibitem [{\citenamefont {Yao}\ and\ \citenamefont {Wang}(2018)}]{YaoWang}%
  \BibitemOpen
  \bibfield  {author} {\bibinfo {author} {\bibfnamefont {S.}~\bibnamefont
  {Yao}}\ and\ \bibinfo {author} {\bibfnamefont {Z.}~\bibnamefont {Wang}},\
  }\bibfield  {title} {\bibinfo {title} {Edge states and topological invariants
  of non-hermitian systems},\ }\href
  {https://doi.org/10.1103/PhysRevLett.121.086803} {\bibfield  {journal}
  {\bibinfo  {journal} {Phys. Rev. Lett.}\ }\textbf {\bibinfo {volume} {121}},\
  \bibinfo {pages} {086803} (\bibinfo {year} {2018})}\BibitemShut {NoStop}%
\bibitem [{\citenamefont {Kunst}\ \emph {et~al.}(2018)\citenamefont {Kunst},
  \citenamefont {Edvardsson}, \citenamefont {Budich},\ and\ \citenamefont
  {Bergholtz}}]{KunstEdvardsson}%
  \BibitemOpen
  \bibfield  {author} {\bibinfo {author} {\bibfnamefont {F.~K.}\ \bibnamefont
  {Kunst}}, \bibinfo {author} {\bibfnamefont {E.}~\bibnamefont {Edvardsson}},
  \bibinfo {author} {\bibfnamefont {J.~C.}\ \bibnamefont {Budich}},\ and\
  \bibinfo {author} {\bibfnamefont {E.~J.}\ \bibnamefont {Bergholtz}},\
  }\bibfield  {title} {\bibinfo {title} {Biorthogonal bulk-boundary
  correspondence in non-hermitian systems},\ }\href
  {https://doi.org/10.1103/PhysRevLett.121.026808} {\bibfield  {journal}
  {\bibinfo  {journal} {Phys. Rev. Lett.}\ }\textbf {\bibinfo {volume} {121}},\
  \bibinfo {pages} {026808} (\bibinfo {year} {2018})}\BibitemShut {NoStop}%
\bibitem [{\citenamefont {Okuma}\ and\ \citenamefont
  {Sato}(2020)}]{NobuyukiSato}%
  \BibitemOpen
  \bibfield  {author} {\bibinfo {author} {\bibfnamefont {N.}~\bibnamefont
  {Okuma}}\ and\ \bibinfo {author} {\bibfnamefont {M.}~\bibnamefont {Sato}},\
  }\bibfield  {title} {\bibinfo {title} {Hermitian zero modes protected by
  nonnormality: Application of pseudospectra},\ }\href
  {https://doi.org/10.1103/PhysRevB.102.014203} {\bibfield  {journal} {\bibinfo
   {journal} {Phys. Rev. B}\ }\textbf {\bibinfo {volume} {102}},\ \bibinfo
  {pages} {014203} (\bibinfo {year} {2020})}\BibitemShut {NoStop}%
\bibitem [{\citenamefont {Ryu}\ \emph {et~al.}(2010)\citenamefont
  {Ryu}, \citenamefont {Schnyder}, \citenamefont {Furusaki},\ and\ \citenamefont
  {Ludwig}}]{RyuSchnyder}%
  \BibitemOpen
  \bibfield  {author} {\bibinfo {author} {\bibfnamefont {S.}~\bibnamefont
  {Ryu}}, \bibinfo {author} {\bibfnamefont {A.~P.}\ \bibnamefont
  {Schnyder}}, \bibinfo {author} {\bibfnamefont {A.}\ \bibnamefont
  {Furusaki}},\ and\ \bibinfo {author} {\bibfnamefont {A.~W.~W.}~\bibnamefont
  {Ludwig}},\ }\bibfield  {title} {\bibinfo {title} {Topological insulators and superconductors: tenfold way and dimensional hierarchy},\ }\href {https://iopscience.iop.org/article/10.1088/1367-2630/12/6/065010}
  {\bibfield  {journal} {\bibinfo  {journal} {New J. Phys.}\ }\textbf {\bibinfo
  {volume} {12}},\ \bibinfo {pages} {065010} (\bibinfo {year}
  {2010})}\BibitemShut {NoStop}%
\bibitem [{\citenamefont {Herviou}\ \emph {et~al.}(2019)\citenamefont
  {Herviou}, \citenamefont {Bardarson},\ and\ \citenamefont
  {Regnault}}]{HerviouBardarson}%
  \BibitemOpen
  \bibfield  {author} {\bibinfo {author} {\bibfnamefont {L.}~\bibnamefont
  {Herviou}}, \bibinfo {author} {\bibfnamefont {J.~H.}\ \bibnamefont
  {Bardarson}},\ and\ \bibinfo {author} {\bibfnamefont {N.}~\bibnamefont
  {Regnault}},\ }\bibfield  {title} {\bibinfo {title} {Defining a bulk-edge
  correspondence for non-hermitian hamiltonians via singular-value
  decomposition},\ }\href {https://doi.org/10.1103/PhysRevA.99.052118}
  {\bibfield  {journal} {\bibinfo  {journal} {Phys. Rev. A}\ }\textbf {\bibinfo
  {volume} {99}},\ \bibinfo {pages} {052118} (\bibinfo {year}
  {2019})}\BibitemShut {NoStop}%
\bibitem{BrunelliNunnenkamp}%
  \BibitemOpen
  \bibfield  {author} {\bibinfo {author} {\bibfnamefont {M.}~\bibnamefont
  {Brunelli}}, \bibinfo {author} {\bibfnamefont {C.~C.}\ \bibnamefont
  {Wanjura}},\ and\ \bibinfo {author} {\bibfnamefont {A.}~\bibnamefont
  {Nunnenkamp}},\ }\bibfield  {title} {\bibinfo {title} {Restoration of the non-Hermitian bulk-boundary correspondence via topological amplification},\ }\href {https://www.scipost.org/SciPostPhys.15.4.173?acad_field_slug=physics}
  {\bibfield  {journal} {\bibinfo  {journal} {SciPost Physics}\ }\textbf {\bibinfo
  {volume} {15}},\ \bibinfo {pages} {173} (\bibinfo {year}
  {2023})}\BibitemShut {NoStop}%
\bibitem{Porras1}%
  \BibitemOpen
  \bibfield  {author} {\bibinfo {author} {\bibfnamefont {D.}~\bibnamefont
  {Porras}}, \ and \ \bibinfo {author} {\bibfnamefont {S.}\ \bibnamefont {Fern\'andez-Lorenzo}},\ } \bibfield  {title} {\bibinfo {title} {Topological amplification in photonic lattices},\ }\href
  {https://link.aps.org/doi/10.1103/PhysRevLett.122.143901} {\bibfield  {journal}
  {\bibinfo  {journal} {Phys. Rev. Lett.}\ }\textbf {\bibinfo {volume} {122}},\
  \bibinfo {pages} {143901} (\bibinfo {year} {2019})}\BibitemShut {NoStop}%
\bibitem{Porras2}%
  \BibitemOpen
  \bibfield  {author} {\bibinfo {author} {\bibfnamefont {T.}~\bibnamefont
  {Ramos}}, \ \bibinfo {author} {\bibfnamefont {J.~J.}\ \bibnamefont {Garc\'{\i}a-Ripoll}} \, \ and \ \bibinfo {author} {\bibfnamefont {D.}\ \bibnamefont {Porras}},\ } \bibfield  {title} {\bibinfo {title} {Topological input-output theory for directional amplification},\ }\href
  {https://link.aps.org/doi/10.1103/PhysRevA.103.033513} {\bibfield  {journal}
  {\bibinfo  {journal} {Phys. Rev. A}\ }\textbf {\bibinfo {volume} {103}},\
  \bibinfo {pages} {033513} (\bibinfo {year} {2021})}\BibitemShut {NoStop}%
  \bibitem [{\citenamefont {Hughes}\ \emph {et~al.}(2011)\citenamefont {Hughes},
  \citenamefont {Prodan},\ and\ \citenamefont {Bernevig}}]{HughesProdan}%
  \BibitemOpen
  \bibfield  {author} {\bibinfo {author} {\bibfnamefont {T.~L.}\ \bibnamefont
  {Hughes}}, \bibinfo {author} {\bibfnamefont {E.}~\bibnamefont {Prodan}},\
  and\ \bibinfo {author} {\bibfnamefont {B.~A.}\ \bibnamefont {Bernevig}},\
  }\bibfield  {title} {\bibinfo {title} {Inversion-symmetric topological
  insulators},\ }\href {https://doi.org/10.1103/PhysRevB.83.245132} {\bibfield
  {journal} {\bibinfo  {journal} {Phys. Rev. B}\ }\textbf {\bibinfo {volume}
  {83}},\ \bibinfo {pages} {245132} (\bibinfo {year} {2011})}\BibitemShut
  {NoStop}%
\bibitem [{\citenamefont {Fang}\ \emph {et~al.}(2013)\citenamefont {Fang},
  \citenamefont {Gilbert},\ and\ \citenamefont {Bernevig}}]{FangGilbert}%
  \BibitemOpen
  \bibfield  {author} {\bibinfo {author} {\bibfnamefont {C.}~\bibnamefont
  {Fang}}, \bibinfo {author} {\bibfnamefont {M.~J.}\ \bibnamefont {Gilbert}},\
  and\ \bibinfo {author} {\bibfnamefont {B.~A.}\ \bibnamefont {Bernevig}},\
  }\bibfield  {title} {\bibinfo {title} {Entanglement spectrum classification
  of ${C}_{n}$-invariant noninteracting topological insulators in two
  dimensions},\ }\href {https://doi.org/10.1103/PhysRevB.87.035119} {\bibfield
  {journal} {\bibinfo  {journal} {Phys. Rev. B}\ }\textbf {\bibinfo {volume}
  {87}},\ \bibinfo {pages} {035119} (\bibinfo {year} {2013})}\BibitemShut
  {NoStop}%
\bibitem [{\citenamefont {Monkman}\ and\ \citenamefont
  {Sirker}(2023{\natexlab{a}})}]{MonkmanSirker3}%
  \BibitemOpen
  \bibfield  {author} {\bibinfo {author} {\bibfnamefont {K.}~\bibnamefont
  {Monkman}}\ and\ \bibinfo {author} {\bibfnamefont {J.}~\bibnamefont
  {Sirker}},\ }\bibfield  {title} {\bibinfo {title} {Symmetry-resolved
  entanglement of ${C}_{2}$-symmetric topological insulators},\ }\href
  {https://doi.org/10.1103/PhysRevB.107.125108} {\bibfield  {journal} {\bibinfo
   {journal} {Phys. Rev. B}\ }\textbf {\bibinfo {volume} {107}},\ \bibinfo
  {pages} {125108} (\bibinfo {year} {2023}{\natexlab{a}})}\BibitemShut
  {NoStop}%
\bibitem [{\citenamefont {Monkman}\ and\ \citenamefont
  {Sirker}(2023{\natexlab{b}})}]{MonkmanSirker4}%
  \BibitemOpen
  \bibfield  {author} {\bibinfo {author} {\bibfnamefont {K.}~\bibnamefont
  {Monkman}}\ and\ \bibinfo {author} {\bibfnamefont {J.}~\bibnamefont
  {Sirker}},\ }\bibfield  {title} {\bibinfo {title} {Symmetry-resolved
  entanglement: general considerations, calculation from correlation functions,
  and bounds for symmetry-protected topological phases},\ }\href
  {https://doi.org/10.1088/1751-8121/ad086d} {\bibfield  {journal} {\bibinfo
  {journal} {Journal of Physics A: Mathematical and Theoretical}\ }\textbf
  {\bibinfo {volume} {56}},\ \bibinfo {pages} {495001} (\bibinfo {year}
  {2023}{\natexlab{b}})}\BibitemShut {NoStop}%
\bibitem [{\citenamefont {Flynn}\ \emph {et~al.}(2021)\citenamefont {Flynn},
  \citenamefont {Cobanera}, \ and\ \citenamefont {Viola}}]{FlynnViola}%
  \BibitemOpen
  \bibfield  {author} {\bibinfo {author} {\bibfnamefont {V.~P.}~\bibnamefont
  {Flynn}}, \bibinfo {author} {\bibfnamefont {E.}\ \bibnamefont {Cobanera}}, \ and \ 
  \bibinfo {author} {\bibfnamefont {L.}\ \bibnamefont {Viola}},\ }\bibfield  {title} {\bibinfo {title} {Topology by dissipation: Majorana bosons in metastable quadratic markovian dynamics},\ }\href
  {https://link.aps.org/doi/10.1103/PhysRevLett.127.245701} {\bibfield  {journal}
  {\bibinfo  {journal} {Phys. Rev. Lett.}\ }\textbf {\bibinfo {volume} {127}},\
  \bibinfo {pages} {245701} (\bibinfo {year} {2021})}\BibitemShut {NoStop}%
\bibitem [{\citenamefont {Flynn2}\ \emph {et~al.}(2023)\citenamefont {Flynn2},
  \citenamefont {Cobanera2}, \ and\ \citenamefont {Viola2}}]{FlynnViola2}%
  \BibitemOpen
  \bibfield  {author} {\bibinfo {author} {\bibfnamefont {V.~P.}~\bibnamefont
  {Flynn}}, \bibinfo {author} {\bibfnamefont {E.}\ \bibnamefont {Cobanera}}, \ and \ 
  \bibinfo {author} {\bibfnamefont {L.}\ \bibnamefont {Viola}},\ }\bibfield  {title} {\bibinfo {title} {Topological zero modes and edge symmetries of metastable Markovian bosonic systems},\ }\href
  {https://link.aps.org/doi/10.1103/PhysRevB.108.214312} {\bibfield  {journal}
  {\bibinfo  {journal} {Phys. Rev. B}\ }\textbf {\bibinfo {volume} {108}},\
  \bibinfo {pages} {214312} (\bibinfo {year} {2023})}\BibitemShut {NoStop}%  
\bibitem [{\citenamefont {Boettcher}\ and\ \citenamefont
  {Sirker}(2023{\natexlab{b}})}]{Boettcher}%
  \BibitemOpen
  \bibfield  {author} {\bibinfo {author} {\bibfnamefont {A.}~\bibnamefont
  {Boettcher}},\ }\bibfield  {title} {\bibinfo {title} {On the approximation numbers of large Toeplitz matrices},\ }\href
  {https://ems.press/journals/dm/articles/8964950} {\bibfield  {journal} {\bibinfo
  {journal} {Doc. Math.}\ }\textbf
  {\bibinfo {volume} {2}},\ \bibinfo {pages} {1} (\bibinfo {year}
  {1997}{\natexlab{b}})}\BibitemShut {NoStop}%
\bibitem [{\citenamefont {{Yin}}\ \emph {et~al.}(2018)\citenamefont {{Yin}},
  \citenamefont {{Jiang}}, \citenamefont {{Li}}, \citenamefont {{L{\"u}}},\
  and\ \citenamefont {{Chen}}}]{YJLLC2018}%
  \BibitemOpen
  \bibfield  {author} {\bibinfo {author} {\bibfnamefont {C.}~\bibnamefont
  {{Yin}}}, \bibinfo {author} {\bibfnamefont {H.}~\bibnamefont {{Jiang}}},
  \bibinfo {author} {\bibfnamefont {L.}~\bibnamefont {{Li}}}, \bibinfo {author}
  {\bibfnamefont {R.}~\bibnamefont {{L{\"u}}}},\ and\ \bibinfo {author}
  {\bibfnamefont {S.}~\bibnamefont {{Chen}}},\ }\bibfield  {title} {\bibinfo
  {title} {{Geometrical meaning of winding number and its characterization of
  topological phases in one-dimensional chiral non-Hermitian systems}},\ }\href
  {https://doi.org/10.1103/PhysRevA.97.052115} {\bibfield  {journal} {\bibinfo
  {journal} {\pra}\ }\textbf {\bibinfo {volume} {97}},\ \bibinfo {eid} {052115}
  (\bibinfo {year} {2018})},\ \BibitemShut {NoStop}%
\bibitem [{\citenamefont {{Jiang}}\ \emph {et~al.}(2018)\citenamefont
  {{Jiang}}, \citenamefont {{Yang}},\ and\ \citenamefont {{Chen}}}]{JYC2018}%
  \BibitemOpen
  \bibfield  {author} {\bibinfo {author} {\bibfnamefont {H.}~\bibnamefont
  {{Jiang}}}, \bibinfo {author} {\bibfnamefont {C.}~\bibnamefont {{Yang}}},\
  and\ \bibinfo {author} {\bibfnamefont {S.}~\bibnamefont {{Chen}}},\
  }\bibfield  {title} {\bibinfo {title} {{Topological invariants and phase
  diagrams for one-dimensional two-band non-Hermitian systems without chiral
  symmetry}},\ }\href {https://doi.org/10.1103/PhysRevA.98.052116} {\bibfield
  {journal} {\bibinfo  {journal} {\pra}\ }\textbf {\bibinfo {volume} {98}},\
  \bibinfo {eid} {052116} (\bibinfo {year} {2018})},\ 
  \BibitemShut {NoStop}%
 \bibitem [{\citenamefont {Naghiloo}\ \emph {et~al.}(2019)\citenamefont {Naghiloo},
  \citenamefont {Wang}, \citenamefont {Rao}, \citenamefont {Gui}, \citenamefont
  {Yao}, \citenamefont {Lu},\ and\ \citenamefont {Hu}}]{NaghilooAbbasi}%
  \BibitemOpen
  \bibfield  {author} {\bibinfo {author} {\bibfnamefont {M.}~\bibnamefont
  {Naghiloo}}, \bibinfo {author} {\bibfnamefont {M.}\ \bibnamefont {Abbasi}},
  \bibinfo {author} {\bibfnamefont {Y.~N.}\ \bibnamefont {Joglekar}}, \ and\ \bibinfo {author} {\bibfnamefont {K.-W.}\
  \bibnamefont {Murch}},\ }\bibfield  {title} {\bibinfo {title} {Quantum state tomography across the exceptional point in a single dissipative qubit},\ }\href
  {https://doi.org/10.1038/s41567-019-0652-z} {\bibfield  {journal}
  {\bibinfo  {journal} {Nat. Phys.}\ }\textbf {\bibinfo {volume} {15}},\
  \bibinfo {pages} {1232} (\bibinfo {year} {2019})}\BibitemShut {NoStop}%
  \bibitem [{\citenamefont
  {Ughrelidze}, \citenamefont {Flynn}, \citenamefont {Cobanera},\ and\
  \citenamefont {Viola}}]{UghrelidzeViola}%
  \BibitemOpen
  \bibfield  {author} {\bibinfo {author} {\bibfnamefont {M.}~\bibnamefont
  {Ughrelidze}}, \bibinfo {author} {\bibfnamefont {V.~P.}~\bibnamefont
  {Flynn}}, \bibinfo {author} {\bibfnamefont {E.}\ \bibnamefont
  {Cobanera}},\ and\ \bibinfo {author} {\bibfnamefont {L.}~\bibnamefont
  {Viola}},\ }
  \bibfield  {title} {\bibinfo {title} {The interplay of finite and infinite size stability in quadratic bosonic lindbladians},}
  \href {https://arxiv.org/abs/2405.08873} {\bibfield
  {journal} {\bibinfo  {journal} {arxiv:2405.08873}\ }\textbf {\bibinfo {volume}
  {}} \bibinfo {pages} {} (\bibinfo {year} {2024})}\BibitemShut
  {NoStop}%  
  \bibitem [{\citenamefont {Monkman}\ and\ \citenamefont
  {Sirker}(2023{\natexlab{c}})}]{MonkmanSirker2}%
  \BibitemOpen
  \bibfield  {author} {\bibinfo {author} {\bibfnamefont {K.}~\bibnamefont
  {Monkman}}\ and\ \bibinfo {author} {\bibfnamefont {J.}~\bibnamefont
  {Sirker}},\ }\bibfield  {title} {\bibinfo {title} {Entanglement and particle
  fluctuations of one-dimensional chiral topological insulators},\ }\href
  {https://doi.org/10.1103/PhysRevB.108.125116} {\bibfield  {journal} {\bibinfo
   {journal} {Phys. Rev. B}\ }\textbf {\bibinfo {volume} {108}},\ \bibinfo
  {pages} {125116} (\bibinfo {year} {2023}{\natexlab{c}})}\BibitemShut
  {NoStop}%
\end{thebibliography}
\end{document}